\documentclass[a4paper]{jpconf}
\usepackage{lipsum} 
\usepackage{tikz}
\usepackage[sc]{mathpazo} 
\usepackage[T1]{fontenc} 
\usepackage{microtype} 
\usepackage{amsmath,mathtools}

\newcommand{\diagentry}[1]{\mathmakebox[1.8em]{#1}}
\newcommand{\xddots}{%
  \raise 4pt \hbox {.}
  \mkern 6mu
  \raise 1pt \hbox {.}
  \mkern 6mu
  \raise -2pt \hbox {.}
}
\usepackage[hmarginratio=1:1, top=32mm, columnsep=20pt]{geometry} 
\usepackage{multicol} 
\usepackage[hang, small, labelfont=bf, up, textfont=it, up]{caption} 
\usepackage{booktabs} 
\usepackage{float} 
\usepackage{hyperref} 
\usepackage{threeparttable}
\usepackage{lettrine} 
\usepackage{paralist} 
\usepackage{graphicx}
\begin{document}
\newgeometry{a4paper,left=25mm,right=25mm,top=40mm,bottom=27mm}
\title{Basis entropy: A useful physical quantity about projective measurement}

\author{Xing Chen}

\address{Institute Of Physics, Chinese Academy Of Sciences, 100190 Beijing, People's Republic of China}

\ead{chenx@iphy.ac.cn}

\begin{abstract}
Projective measurement can increase the entropy of a state $\rho$, the increased entropy is not only up to the basis of projective measurement, but also has something to do with the properties of the state itself. In this paper we define this increased entropy as basis entropy. And then we discuss the usefulness of this new concept by showing its application in deciding whether a state is pure or not and detecting the existence of quantum discord. And as shown in the paper, this new concept can also be used to describe decoherence.
\end{abstract}

\section{Introduction}
Projective measurement can increase the entropy of a state $\rho$\cite{nie}\cite{ben}. And for different states, the increased entropy is different, and it is up to two factors
\begin{enumerate}
	 \item The orthogonal projectors of projective measurement
  \item The state itself.
\end{enumerate}
\indent Every state has its increased entropy after a projective measurement. This increased entropy is actually a quite useful physical quantity. The aim of this paper is to show the usefulness of this increased entropy by providing some theorems about this increased entropy. A mentionable merit of this increased entropy is that it is highly related to quantum discord, and it can be used to detect the existence of quantum discord.\\

\section{Basis entropy}
\begin{figure}[H]
  		\centering
 		 \includegraphics[height=180pt,width=270pt]{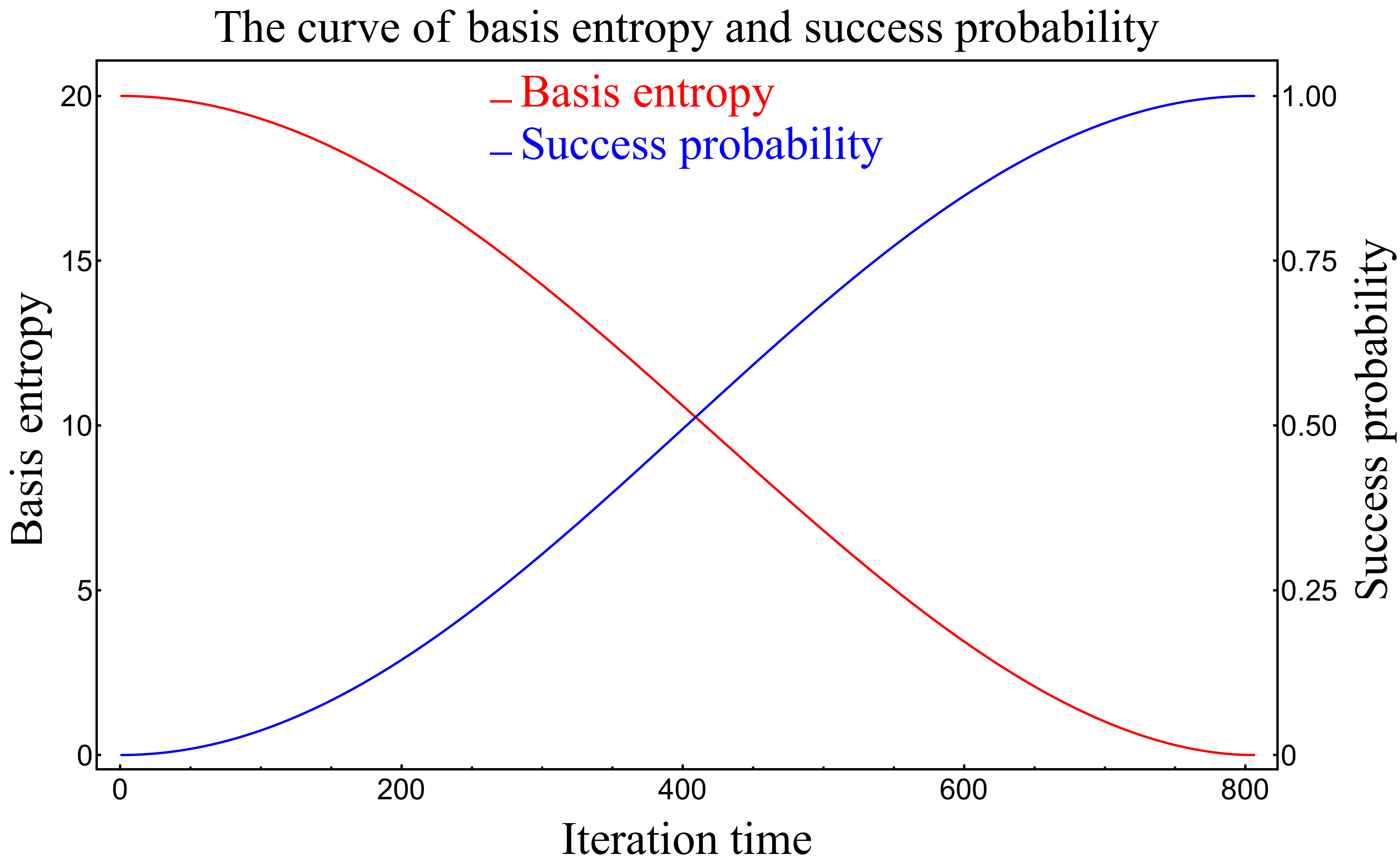}
	\caption{Basis entropy and success probability change of Grover's algorithm}\label{fig1}
\end{figure}

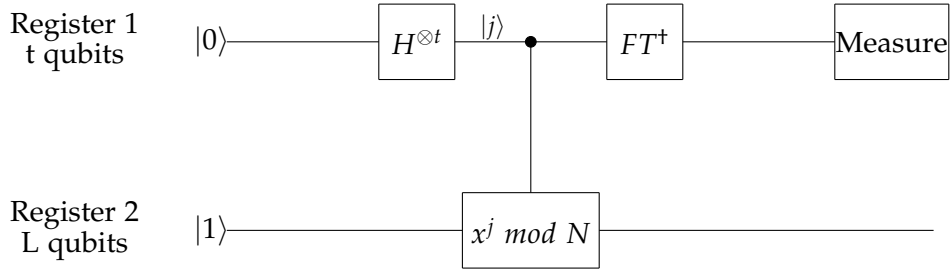
\begin{figure}[H]
\centering
  \begin{tikzpicture}
\draw (0,0) -- (1,0);
\draw (1,0) -- (1,1);
\draw (1,1) -- (0,1);
\draw (0,1) -- (0,0);
\node at (0.5,0.5) {$H^{\otimes t}$};
\draw (1,0.5) -- (3,0.5);
\node at (1.5,0.7) {\footnotesize$|j\rangle$};
\draw (3,0) -- (4,0);
\draw (4,0) -- (4,1);
\draw (4,1) -- (3,1);
\draw (3,1) -- (3,0);
\node at (3.5,0.5) {$FT^{\dag}$};
\draw (4,0.5) -- (6,0.5);
\draw (6,0) -- (7.5,0);
\draw (7.5,0) -- (7.5,1);
\draw (7.5,1) -- (6,1);
\draw (6,1) -- (6,0);
\node at (6.75,0.5) {Measure};
\draw (2,0.5) -- (2,-1.5);
\draw (1.1,-2.5) -- (2.9,-2.5);
\draw (2.9,-2.5) -- (2.9,-1.5);
\draw (2.9,-1.5) -- (1.1,-1.5);
\draw (1.1,-1.5) -- (1.1,-2.5);
\node at (2,-2) {$x^j\ mod\ N$};
\draw (-2,0.5) -- (0,0.5);
\draw (-2,-2) -- (1.1,-2);
\draw (2.9,-2) -- (7.3,-2);
\filldraw[black] (2,0.5) circle (2pt);
\node at (-2.2,0.5) {$|0\rangle$};
\node at (-2.2,-2) {$|1\rangle$};
\node at (-4,0.7) {Register 1};
\node at (-4,0.3) {t qubits};
\node at (-4,-1.8) {Register 2};
\node at (-4,-2.2) {L qubits};
\end{tikzpicture}
\caption{Quantum circuit of Shor's algorithm}\label{fig2}
\end{figure}

First we should give this increased entropy a proper name. Since projective measurement is dependent on its basis, we suggest using 'basis entropy' to describe this increased entropy of the state. The meaning of which is, with the knowledge of the state(the knowledge here means the information which equals to the von Neumann entropy of this state), the ignorance of a state's projective measurement result. And according to the definition of basis entropy, we can use the following formula to calculate basis entropy
\begin{equation}
\emph{BE}=S(\sum_iP_i\rho P_i)-S(\rho)
\end{equation}
where $P_i$ is a complete set of orthogonal projectors.\\
\indent A good example which can illustrate the physical meaning of basis entropy is Grove's algorithm\cite{gro}. First let us review the procedure of Grover's algorithm\cite{nie}.
\begin{enumerate}
\item $|0\rangle^{\otimes n}|1\rangle$
\item $\longrightarrow\frac{1}{\sqrt{2^n}}\sum_{x=0}^{2^n-1}|x\rangle[\frac{|0\rangle-|1\rangle}{\sqrt{2}}] $
\item $\longrightarrow[(2|\psi\rangle\langle\psi|-\textbf{I})\textbf{O}]^R\frac{1}{\sqrt{2^n}}\sum_{x=0}^{2^n-1}|x\rangle[\frac{|0\rangle-|1\rangle}{\sqrt{2}}]\\
    \approx|x_0\rangle[\frac{|0\rangle-|1\rangle}{\sqrt{2}}]$
\item $\longrightarrow x_0$
\end{enumerate}
The first state is the initial state, where $n$ represents the qubit number, by applying $H^{\otimes n}$ to initial $n$ qubits we get the state in step two, and then we apply the Grover iteration $(2|\psi\rangle\langle\psi|-\textbf{I})\textbf{O}$\cite{note1}, we get the state in step three. After applying the Grover iteration for about $R\approx\lceil\pi\sqrt{2^n}/4\rceil$ times, we get the wanted state $|x_0\rangle$, the detail of Grover's algorithm refer \cite{nie}.\\
\indent For Grover's algorithm, we actually search in the state $(\sum_{i=0}^{2^n-1}|x_i\rangle)/\sqrt{2^n}$, and measure it with projectors $\{|0\rangle\langle0|,|1\rangle\langle1|\}$, before apply Grover iteration, the basis entropy of the database state is $n$, this is our ignorance of the measurement result, which means without applying Grover iteration, we need $n$ bits to describe the measurement result. After apply one time of Grover iteration, the basis entropy decreased, which means our ignorance of the measurement result decreased, so the probability of finding the target state $|x_0\rangle$ increased. As shown in Figure \ref{fig1}, with the decreasing of the basis entropy, the successful probability is increasing.\\
\indent Figure \ref{fig1} is the basis entropy change of Grover's algorithm, here we set $n=20$, clearly we can see from this figure that, the more time we apply Grover iteration(under the desired time $R\approx\lceil\pi\sqrt{2^n}/4\rceil$, see appendix A), the basis entropy will be smaller and the success probability will be higher. \\
Another example is Shor's algorithm. The procedure of Shor's algorithm is shown as follows \cite{nie}:
\begin{multicols}{2}
\begin{enumerate}
	\item $|0\rangle|1\rangle$\\
  \item $\longrightarrow\frac{1}{\sqrt{2^t}}\sum_{j=0}^{2^t-1}|j\rangle|1\rangle$
\end{enumerate}
   initial state\\ \quad\\
  create superposition
\end{multicols}
\begin{multicols}{2}
\begin{enumerate}
 \setcounter{enumi}{2}
    \item $\longrightarrow\frac{1}{\sqrt{2^t}}\sum_{j=0}^{2^t-1}|j\rangle|x^j mod N\rangle\\
  			\approx\frac{1}{r2^t}\sum_{s=0}^{r-1}\sum_{j=0}^{2^t-1}e^{2\pi isj/r}|j\rangle|u_s\rangle$\\
  \item $\longrightarrow\frac{1}{\sqrt{r}}\sum_{s=0}^{r-1}|\widetilde{s/r}\rangle|u_s\rangle$\\
  \item $\longrightarrow\widetilde{s/r}$\\
  \item $\longrightarrow r$\\
\end{enumerate}
   apply $U_{x,N}$\\ \quad\\ \quad\\
  apply inverse Fourier transform to first register\\ \quad\\
   measure first register\\ \quad\\
   apply continued fractions algorithm
\end{multicols}
where $U_{x,N}$ performs the transformation $|j\rangle|k\rangle\longrightarrow|j\rangle|x^jk$ $mod N\rangle$. $r$ is the least integer such that $x^r=1(mod N)$, Figure \ref{fig2} is the quantum circuit of Shor's algorithm.
where $t$ is the qubit number of $|0\rangle$, and $L$ is the qubit number of $|1\rangle$. For more details of Shor's algorithm, refer \cite{nie}. Next, we will analyse the basis entropy of Shor's algorithm.\\
\indent In Shor's algorithm, we use state $(\sum_{j=0}^{2^t-1}|j\rangle|1\rangle)/\sqrt{2^t}$ to factor great numbers, but we get the useful measurement result from the first register, so we only calculate the basis entropy of the first register. The measurement projectors of Shor's algorithm are $\{|0\rangle\langle0|,|1\rangle\langle1|\}$. Before we apply Shor's algorithm, the basis entropy of the first register is $t$, and then after the $U_{x,N}$ transformation, the basis entropy remains unchanged, and after the inverse Fourier transform, the basis entropy becomes $r$, then the main part of Shor's algorithm is over, the subsequent steps are measuring the state and applying mathematical methods to the measurement result. \\
\indent From the change of the basis entropy we can easily see that the most crucial step of Shor's algorithm is applying inverse Fourier transform, because this step reduced the basis entropy of the first register without decohering it. After this crucial step, the basis entropy of the first register reduced from $t$ to $r$, and then we have a higher probability to get the desired state by measuring the first register.\\
\indent From the two examples above we can see that basis entropy is the ignorance of the measurement result with the knowledge of the state.
\section{Maximum and minimum basis entropy}
A state's basis entropy is dependent on different measurement projectors, so the number of its basis entropy is actually infinite, but we are only interested in its maximum and minimum basis entropy.
\subsection{The property of maximum basis entropy}
For a state's maximum basis entropy, we have the following theorem.
\begin{quote}
  Theorem 1. Iff a state's maximal basis entropy is $log_2D$, the state is a pure state.
\end{quote}
where $D$ is the dimension of Hilbert space.\\
\indent Proof: If a state's maximal basis entropy is $log_2D$, which means
\begin{equation}
\emph{BE}_{max}=S(\sum_iP_i\rho P_i)-S(\rho)=log_2D
\end{equation}
Since the entropy of a state is non-negative and in a D-dimensional Hilbert space, the state's von Neumann entropy is at most $log_2D$\cite{nie}, so if the basis entropy of a state is $log_2D$, then $S(\rho)$ must be zero, which means the state is a pure state.\\
Let's prove the other direction of the theorem. It seems only for pure state like
\begin{equation}
|\psi\rangle=\frac{1}{\sqrt{D}}\sum_i^D|i\rangle
\end{equation}
has $log_2D$ basis entropy, while for pure state like $(\sqrt{3}|0\rangle+|1\rangle)/2$, its basis entropy is smaller than $log_2D$. We will prove that for any pure state, its maximum basis entropy will be $log_2D$, as long as we choose right projectors to measure them. For simplicity, we just prove the case $D=2$. For any pure state, we can write its density matrix as
\begin{equation}
  \rho=\frac12\textbf{I}+a\textbf{$\sigma_1$}+b\textbf{$\sigma_2$}+c\textbf{$\sigma_3$}
\end{equation}
where $a,b$ and $c$ are coefficients and $\sigma_1,\sigma_2,\sigma_3$ are Pauli matrices. We only need to prove that
\begin{equation}
S_{max}(\sum_i^D P_i\rho P_i)=1
\end{equation}
where,
\begin{equation}
\{P_k=V|k\rangle\langle k|V^{\dagger}:k=0,1\}
\end{equation}
is the complete set of orthogonal projectors. And $V$ is a 2-dimensional unitary transformation. It has be proven that for any 2-dimensional pure state, its maximum basis entropy is $log_2 2=1$ (for those who are interested in the details ,see appendix B). And now we complete the proof of theorem 1. The proof can be easily extended to higher Hilbert space.\\
\indent From theorem 1, it is natural to get the following corollary:
\begin{quote}
  Corollary 1. Only for states like $\rho=\sum_i^D |i\rangle\langle i|/D$, they have no basis entropy, it means no projective measurement can increase their entropy.
\end{quote}
This corollary is easy to prove, since for a state like $\rho=\sum_i^D |i\rangle\langle i|/D$, its von Neumann has reached the maximal value in its Hilbert space, so no projective measurement can increase its entropy.\\
\indent We can use maximum basis entropy to judge whether a state is mixed or not. According to theorem 1, if a state's maximum basis entropy is $log_2D$, then the state is a pure state. If its maximum basis entropy is smaller than $log_2D$, then it's a mixed state. If its maximum basis entropy is zero, it has no basis entropy, then the state is $\rho=\sum_i^D |i\rangle\langle i|/D$.\\
\subsection{The property of minimum basis entropy}
For most states, regardless of pure states or mixed states, their minimum basis entropy is zero, which means there exists a complete set of orthogonal projectors, under these projectors, we can get the full knowledge of this state. But there are some states, their minimum basis entropy are not zero, which means no matter what projectors we use to measure them, there is always some information we can not get. This inaccessible information, we will show next, is related to quantum discord.\\
\indent Quantum discord comes from the projective measurement\cite{note2}, just as the basis entropy. For a bell state, its basis entropy is a constant, so it's easy for us to calculate its minimum basis entropy. So let's take a bell state as an example to explain why the quantum discord is related to minimum basis entropy. For bell state like $(|00\rangle+|11\rangle)/\sqrt{2}$, the subsystem is $A$ and $B$, the discord of this state is
\begin{equation}
\begin{split}
  \delta(A:B)_{\{\Pi_i^B\}}=&I(A:B)-J(A:B)_{\{\Pi_i^B\}}\\
  =&S(A)+S(B)-S(A,B)-S(A)+S(A|{\Pi_i^B})\\
  =&S(B)+S(A|{\Pi_i^B})\\
  =&S(B)\\
  =&1
\end{split}
\end{equation}
where $I(A:B)$ is the quantum mutual information between $A$ and $B$, and $J(A:B)_{\{\Pi_i^B\}}$ is the measurement mutual information between $A$ and $B$. From this equation we can see that the quantum discord of this state comes from $S(B)$. Then let's calculate the basis entropy of the joint system $AB$, measure $AB$ by projectors $\{|00\rangle\langle00|,|01\rangle\langle01|,|10\rangle\langle10|,|11\rangle\langle11|\}$, and after this projective measurement the whole system $AB$ will become
 \begin{equation}
   \rho_{AB_{pm}}=\frac12
   \begin{pmatrix}
     1&0&0&0\\
     0&0&0&0\\
     0&0&0&0\\
     0&0&0&1
   \end{pmatrix}
 \end{equation}
 and
 \begin{equation}
   \emph{BE}_{\beta_{00}}=S(\rho_{AB_{pm}})-S(\rho_{AB})=1=S(B)
 \end{equation}
Therefore the basis entropy of the joint system $AB$ is equal to the quantum discord between its subsystems.\\
\begin{figure}[!ht]
  		\centering
 		 \includegraphics[height=180pt,width=270pt]{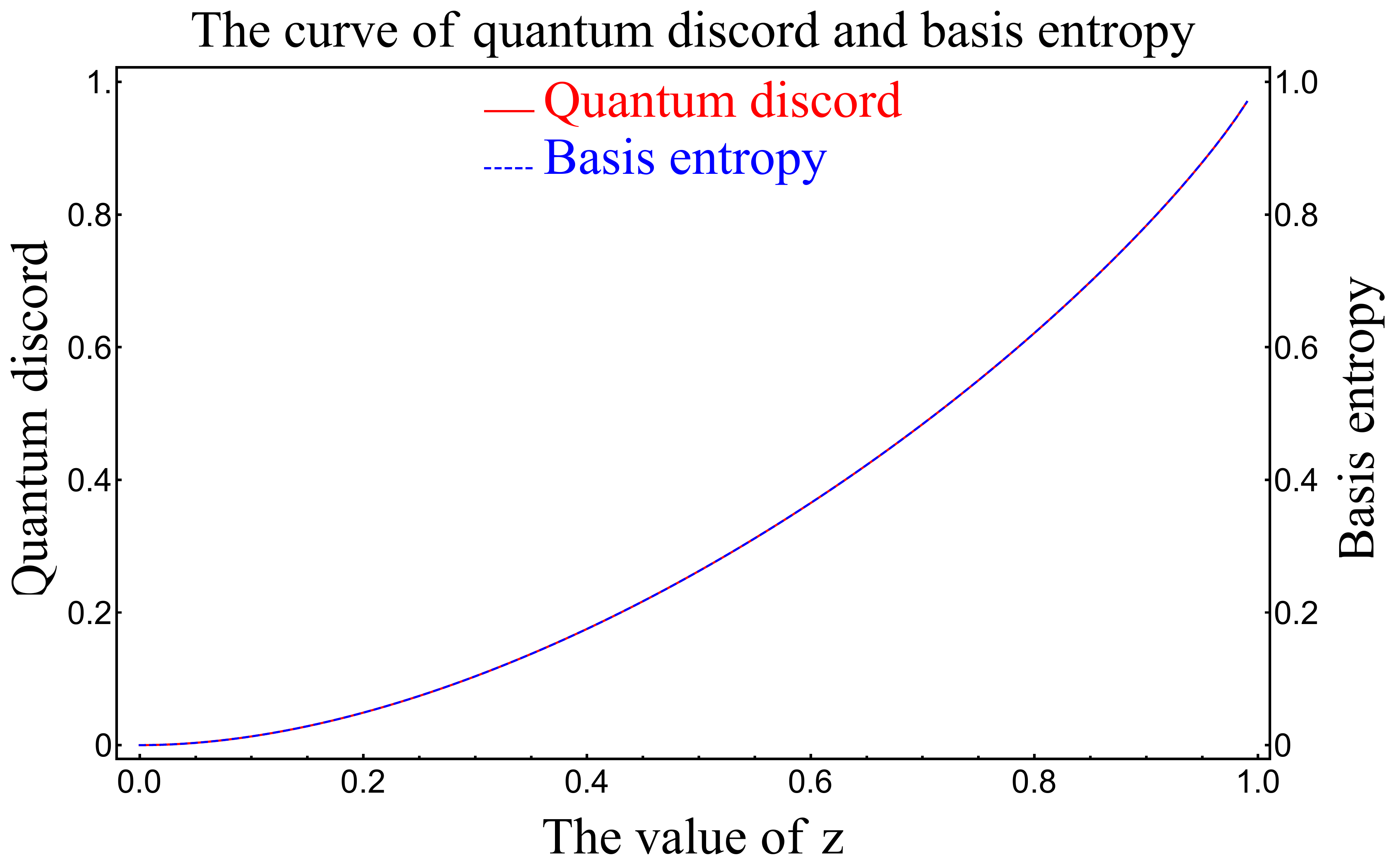}
	\caption{Quantum discord and basis entropy of Werner state}\label{fig3}
\end{figure}
\indent For another example, let's consider the discord of Werner state\cite{oll} $\rho=(1-z)\textbf{I}/4+z|\psi\rangle\langle\psi|$ with $|\psi\rangle=(|00\rangle+|11\rangle)/\sqrt{2}$. One can calculate that the quantum discord of this Werner state is equal to its basis entropy. For example as shown in Figure \ref{fig3}, when $z=1/3$, the value of quantum discord is $0.1258$, it's basis entropy is also $0.1258$.\\
\indent The quantum discord of Bell states or Werner state are exactly equal to their basis entropy, no matter which orthogonal projectors we used to measure them, like $\{|0\rangle\langle0|,|1\rangle\langle1|\}$ or $\{|+\rangle\langle+|,|-\rangle\langle-|\}$. This equivalence has two reasons. First, Bell states or Werner state are invariant under local rotations, so their basis entropy is invariant under different orthogonal projectors; second, for bell state or werner states, they are states with the maximally mixed marginals, thus their quantum discord is equal to their minimum basis entropy. This is also true for ordinary two-qubit system $\rho=(\textbf{I}+\sum_{i=1}^3c_i\sigma_i\otimes\sigma_i)/4$\cite{luo} with maximum mixed marginals, its basis entropy is not a constant. We have shown in appendix C that, its quantum discord is still equal to its minimum basis entropy. For states without maximum marginals, its nonzero minimum basis entropy can guarantee its nonzero quantum discord, so we have the following theorem\\
\begin{quote}
  Theorem 2. If a state's minimum basis entropy is nonzero, there must be quantum discord between its subsystems.
\end{quote}
From the above example we can see that, quantum discord is a symmetric physical quantity for states like $\rho=(\textbf{I}+\sum_{i=1}^3c_i\sigma_i\otimes\sigma_i)/4$. For asymmetric quantum discord, the theorem still holds, see discussion in appendix C\\

\section{Basis entropy and coherence}
One can see that basis entropy is also highly related to quantum coherence\cite{bau}. Except the formula we use to measure basis entropy showed up naturally. One can find that the formula use to quantify coherence is actually a special case of basis entropy, it's a basis entropy with fixed basis, which means under a specific basis, there is no difference between basis entropy and coherence. But we can easily see that if a state has no coherence, we can't guarantee its maximum basis entropy is zero; if a state's maximum basis entropy is zero, then it must have no coherence. This difference between these two concepts has physical meaning. For example for state $|0\rangle$, it has no coherence, we can't decohere it, since its density matrix has no off-diagonal elements. But this state's maximum basis entropy is 1, from this perspective, we can still further decohere it. \\
\indent The decoherence about coherence is basis-dependent\cite{sch}, which means the state can only be seen as decohered under specific basis. The decoherence about maximum basis entropy is not basis-dependent, which means if we decohere a state's maximum basis entropy, the state will be fully decohered and it can't be further decohered. Take the following state as an example.
\begin{equation}
  \rho_0=
  \begin{pmatrix}
    3/4&\sqrt{3}/2\\
    \sqrt{3}/2&1/4
  \end{pmatrix}
\end{equation}
We decohere this state under basis $\{|0\rangle,|1\rangle\}$, it will become
\begin{equation}
  \rho'_0=
  \begin{pmatrix}
    3/4&0\\
    0&1/4
  \end{pmatrix}
\end{equation}
According to corollary 1, this state has basis entropy, so it can be further decohered. The corresponding projectors are
\begin{equation}
P_1=\frac12
  \begin{pmatrix}
    1&-i\\
    i&1
  \end{pmatrix}
  ,P_2=\frac12
    \begin{pmatrix}
     1&i\\
     -i&1
  \end{pmatrix}
\end{equation}
After this , the state becomes
\begin{equation}
  \rho''_0=
  \begin{pmatrix}
    1/2&0\\
    0&1/2
  \end{pmatrix}
\end{equation}
According to corollary 1, this state has no basis entropy, so it can't be further decohered.\\
\indent Therefore the maximum basis entropy can be used as a criterion to measure how much the environment can maximum decohere a quantum system, for pure states, they are highly isolated states and they have the maximum basis entropy, so they are easy to be decohered. And for state like $\rho=\sum_i^D |i\rangle\langle i|/D$, is actually a classical state, its maximum basis entropy is zero, so it has been fully decohered, the environment can't further decohere it.\\

\section{Conclusions}
In this paper we have shown that, the increased entropy of a state after a projective measurement is a useful physical quantity, we named this increased entropy as basis entropy and then we showed some usefulness of this new concept. The maximum basis entropy can be used to decide whether a state is pure or not, and the minimum basis entropy can be used to detect the existence of quantum discord. At the end of this paper, we also discussed the difference between basis entropy and coherence, and showed that basis entropy can be used to describe decoherence.\\
\ack This research was supported by NSF of China, Grant No.11475254. And the help from Qikai He was gratefully acknowledged.
\appendix
\section*{Appendix A}
\setcounter{section}{1}
\indent The success probability of Grover's algorithm: The initial search state is $\frac{1}{\sqrt{2^n}}\sum_{x=0}^{2^n-1}|x\rangle$, the target state is $|x_0\rangle$, we can rewrite the initial state as the combination of target state and un-target states. First of all let's define two normalized states
\begin{equation}
\begin{split}
|\alpha\rangle\equiv&\frac{1}{\sqrt{2^n-1}}\sum_{\substack{x=0 \\ x\neq x_0}}^{2^n-1}|x\rangle\\
|\beta\rangle\equiv&|x_0\rangle
\end{split}
\end{equation}
where $|\alpha\rangle$ represents a sum over all $x$ which are not the state we're searching for, $|\beta\rangle$ represents the target state $|x_0\rangle$.
Then the initial search state can be write
\begin{equation}
\begin{split}
|\psi\rangle=&\frac{1}{\sqrt{2^n}}\sum_{x=0}^{2^n-1}|x\rangle\\
=&\sqrt{\frac{2^n-1}{2^n}}|\alpha\rangle+\frac{1}{\sqrt{2^n}}|\beta\rangle
\end{split}
\end{equation}
As mentioned in \cite{nie}, we can define $cos(\theta/2)=\sqrt{(2^n-1)/2^n}$, then $|\psi\rangle$ can be write as $|\psi\rangle=cos(\theta/2)|\alpha\rangle+sin(\theta/2)|\beta\rangle$. Then after applying the Grover iteration $G=(2|\psi\rangle\langle\psi|-\textbf{I})\textbf{O}$\cite{nie} the state becomes
\begin{equation}
G|\psi\rangle=cos\frac{3\theta}{2}|\alpha\rangle+sin\frac{3\theta}{2}|\beta\rangle.
\end{equation}
After applying $G$ for $k$ times, the state becomes
\begin{equation}
G^k|\psi\rangle=cos\frac{2k+1}{2}\theta|\alpha\rangle+sin\frac{2k+1}{2}\theta|\beta\rangle.
\end{equation}
Then the success probability is the square of the amplitude of $|\beta\rangle$
\begin{equation}
p_{success}=sin^2[\frac{2k+1}{2}\theta]
\end{equation}
The probability $p_{success}$ reaches its maximum when $k=k_{max}$, $k_{max}$ should satisfy the following condition:
\begin{equation}
(2k_{max}+1)\frac{\theta}{2}=\frac{\pi}{2}
\end{equation}
considering $k$ must be an integer, so we get $k_{max}=\lceil\frac{\pi}{2\theta}-\frac12\rceil$. Generally, $2^n$ is very large and thus $\theta$ is very small, so the following approximation is acceptable
\begin{equation}
\theta\approx sin\theta\approx=\frac{2}{\sqrt{2^n}}
\end{equation}
so $k_{max}\approx\lceil\pi\sqrt{2^n}/4\rceil$, this is the desired time we mentioned in the paper before. It is obvious that under the desired time, with the increase of $k$, the success probability $p_{success}$ is increasing. For $n=20$, the desired time is approximately $805$, as shown in Figure \ref{fig1}.\\
\indent The basis entropy of Grover's algorithm: Since we're searching in the state $|\psi\rangle=(\sum_{i=0}^{2^n-1}|x_i\rangle)/\sqrt{2^n}$, so we begin with this state. It's not difficult to check that under measurement projectors $\{|0\rangle\langle0|,|1\rangle\langle1|\}$, its basis entropy is $n$. No matter how many times we apply the Grover iteration, the state $G^k|\psi\rangle$ is a pure state, its von Neumann entropy is zero, so we just need to calculate the entropy of the state after projective measurement, which is
\begin{equation}
  \rho_{pm}=|0\rangle\langle0|G^k|\psi\rangle\langle\psi|G^k|0\rangle\langle0|+|1\rangle\langle1|G^k|\psi\rangle\langle\psi|G^k|1\rangle\langle1|
\end{equation}
which can be write as
\[
\rho_{pm}=
\begin{pmatrix}
\diagentry{ }\\
&\diagentry{cos^2\frac{2k+1}{2}/(2^n-1)}\\
&&\diagentry{\xddots}\\
&&&\diagentry{cos^2\frac{2k+1}{2}/(2^n-1)}\\
&&&&\diagentry{sin^2\frac{2k+1}{2}/2^n}\\
&&&&&\diagentry{cos^2\frac{2k+1}{2}/(2^n-1)}\\
&&&&&&\diagentry{\xddots}\\
&&&&&&&\diagentry{cos^2\frac{2k+1}{2}/(2^n-1)}\\
&&&&&&&&\diagentry{ }\\
\end{pmatrix}_{2^n\times2^n}
\]
so we get
\begin{equation}
S(\rho_{pm})=-[cos^2\frac{2k+1}{2}log_2(cos^2\frac{2k+1}{2}/(2^n-1))+sin^2\frac{2k+1}{2}/2^nlog_2(sin^2\frac{2k+1}{2}/2^n)]
\end{equation}
Thus
\begin{equation}
\begin{split}
\emph{BE}_{grover}=&S(\rho_{pm})-S(G^k|\psi\rangle\langle\psi|G^k)\\
=&-[cos^2\frac{2k+1}{2}log_2(cos^2\frac{2k+1}{2}/(2^n-1))+sin^2\frac{2k+1}{2}/2^nlog_2(sin^2\frac{2k+1}{2}/2^n)]
\end{split}
\end{equation}
With the increase of $k$, the basis entropy of the state $G^k|\psi\rangle$ is decreasing, while the success probability is increasing, as shown in Figure \ref{fig1}.
\appendix
\section*{Appendix B}
\setcounter{section}{2}
In this appendix, we will prove that for any 2-dimensional pure state, its maximum basis entropy is 1. For any 2-dimensional pure state, we can write its density matrix as
\begin{equation}
\rho=\frac12\textbf{I}+a\textbf{$\sigma_1$}+b\textbf{$\sigma_2$}+c\textbf{$\sigma_3$}
\end{equation}
where $a,b,c$ are coefficients and $\sigma_1,\sigma_2,\sigma_3$ are pauli matrices. The measurement projectors are
\begin{equation}
\{B_k=V\Pi_kV^{\dagger}:k=0,1\}
\end{equation}
where $\{\Pi_k=|k\rangle\langle k|:k=0,1\}$, $V\in U(2)$. And $V$ can be written as $V=t\textbf{I}+i\overrightarrow{y}\overrightarrow{\sigma}$, with $t\in R$, $\overrightarrow{y}=(y_1,y_2,y_3)\in R^3$, and $t^2+y_1^2+y_2^2+y_3^2=1$. We only need to prove that
\begin{equation}
S(B_1\rho B_1+B_2\rho B_2)=1.
\end{equation}
After calculation we get
\begin{equation}
\rho_{pm}=B_1\rho B_1+B_2\rho B_2=\frac12\textbf{I}+(az_1+bz_2+cz_3)(z_1\sigma_1+z_2\sigma_2+z_3\sigma_3)
\end{equation}
where
\begin{equation}
\begin{split}
z_1=2(-ty_2+y_1y_3);\\
z_2=2(ty_1+y_2y_3);\\
z_3=t^2+y_3^2-y_1^2-y_2^2.
\end{split}
\end{equation}
After further calculation, we get the eigenvalues of the $\rho_{pm}$
\begin{equation}
\begin{split}
\lambda_1=\frac12-\sqrt{}(a^2 z_1^4+a^2 z_2^2 z_1^2+a^2 z_3^2 z_1^2+2 a b z_2 z_1^3+2 a b z_2^3 z_1+2 a b z_2 z_3^2 z_1+2 a c z_3 z_1^3+2 a c z_3^3 z_1+2 a c z_2^2 z_3 z_1\\
+b^2 z_2^2 z_1^2+b^2 z_2^4+b^2 z_2^2 z_3^2+2 b c z_2 z_3 z_1^2+2 b c z_2 z_3^3+2 b c z_2^3 z_3+c^2 z_3^2 z_1^2+c^2 z_3^4+c^2 z_2^2 z_3^2);\\
\lambda_2=\frac12+\sqrt{}(a^2 z_1^4+a^2 z_2^2 z_1^2+a^2 z_3^2 z_1^2+2 a b z_2 z_1^3+2 a b z_2^3 z_1+2 a b z_2 z_3^2 z_1+2 a c z_3 z_1^3+2 a c z_3^3 z_1+2 a c z_2^2 z_3 z_1\\
+b^2 z_2^2 z_1^2+b^2 z_2^4+b^2 z_2^2 z_3^2+2 b c z_2 z_3 z_1^2+2 b c z_2 z_3^3+2 b c z_2^3 z_3+c^2 z_3^2 z_1^2+c^2 z_3^4+c^2 z_2^2 z_3^2)
\end{split}
\end{equation}
In order to make $S(\rho_{pm})=1$, we need to make sure
\begin{equation}
\begin{split}
\sqrt{}(a^2 z_1^4+a^2 z_2^2 z_1^2+a^2 z_3^2 z_1^2+2 a b z_2 z_1^3+2 a b z_2^3 z_1+2 a b z_2 z_3^2 z_1+2 a c z_3 z_1^3+2 a c z_3^3 z_1+2 a c z_2^2 z_3 z_1\\
+b^2 z_2^2 z_1^2+b^2 z_2^4+b^2 z_2^2 z_3^2+2 b c z_2 z_3 z_1^2+2 b c z_2 z_3^3+2 b c z_2^3 z_3+c^2 z_3^2 z_1^2+c^2 z_3^4+c^2 z_2^2 z_3^2)=0.
\end{split}
\end{equation}
 Using (B.5) and the following condition
\begin{equation}
t^2+y_1^2+y_2^2+y_3^2=1
\end{equation}
we get formula (B.7) equals $2a(-ty_2+y_1y_3)+2b(ty_1+y_2y_3)+c(t^2+y_3^2-y_1^2-y_2^2)$, then we only need to make sure $2a(-ty_2+y_1y_3)+2b(ty_1+y_2y_3)+c(t^2+y_3^2-y_1^2-y_2^2)=0$. Clearly, this equation has qualified solutions, for example, one of the solutions is
\begin{equation}
\begin{split}
&t=\frac{1}{a c}\left(2 a^2 \sqrt{\frac{a^2}{P}+\frac{c^2}{P}-\frac{b c}{P}-\frac{\sqrt{a^4-2 a^2 b c}}{P}}-\right.
b c \sqrt{\frac{a^2}{P}+\frac{c^2}{P}-\frac{b c}{P}-\frac{\sqrt{a^4-2 a^2 b c}}{P}}+\\
&c^2 \sqrt{\frac{a^2}{P}+\frac{c^2}{P}-\frac{b c}{P}-\frac{\sqrt{a^4-2 a^2 b c}}{P}}-
4 a^2 \left(\frac{a^2}{P}+\frac{c^2}{P}-\frac{b c}{P}-\frac{\sqrt{a^4-2 a^2 b c}}{P}\right)^{3/2}-\\
&2 b^2 \left(\frac{a^2}{P}+\frac{c^2}{P}-\frac{b c}{P}-\frac{\sqrt{a^4-2 a^2 b c}}{P}\right)^{3/2}+
4 b c \left(\frac{a^2}{P}+\frac{c^2}{P}-\frac{b c}{P}-\frac{\sqrt{a^4-2 a^2 b c}}{P}\right)^{3/2}-\\
&\left.2 c^2 \left(\frac{a^2}{P}+\frac{c^2}{P}-\frac{b c}{P}-\frac{\sqrt{a^4-2 a^2 b c}}{P}\right)^{3/2}\right);\\
&y_1=0;\\
&y_2=\sqrt{\frac{a^2}{P}+\frac{c^2}{P}-\frac{b c}{P}-\frac{\sqrt{a^4-2 a^2 b c}}{P}};\\
&y_3=\sqrt{\frac{a^2}{P}+\frac{c^2}{P}-\frac{b c}{P}-\frac{\sqrt{a^4-2 a^2 b c}}{P}}.[
\end{split}
\end{equation}
where $P=2(2 a^2+b^2-2 b c+c^2)$
And then we can get
\begin{equation}
V=
\begin{pmatrix}
t+iy_3&iy_1+y_2\\
iy_1-y_2&t-iy_3
\end{pmatrix}
\end{equation}
Then measure $\rho=\frac12\textbf{I}+a\textbf{$\sigma_1$}+b\textbf{$\sigma_2$}+c\textbf{$\sigma_3$}$ by projectors $\{B_k=V\Pi_kV^{\dagger}:k=0,1\}$, we can get $S(B_1\rho B_1+B_2\rho B_2)=1$
\appendix
\section*{Appendix C}
\setcounter{section}{3}
For an ordinary two-qubit state like
\begin{equation}
\begin{split}
\rho=&\frac14(\textbf{I}+\sum_{i=1}^3c_i\sigma_i\otimes\sigma_i)\\
=&\frac14
\begin{pmatrix}
1+c_3&0&0&c_1-c_2\\
0&1-c_3&c_1+c_2&0\\
0&c_1+c_2&1-c_3&0\\
c_1-c_2&0&0&1+c_3
\end{pmatrix}
\end{split}
\end{equation}
As shown in \cite{luo}, the quantum discord of this state is
\begin{equation}
\begin{split}
\emph{Q}=&\emph{I}(\rho)-\emph{C}(\rho)\\
=&\frac14[(1-c_1-c_2-c_3)log_2(1-c_1-c_2-c_3)\\+
&(1-c_1+c_2+c_3)log_2(1-c_1+c_2+c_3)\\+
&(1+c_1-c_2+c_3)log_2(1+c_1-c_2+c_3)\\+
&(1+c_1+c_2-c_3)log_2(1+c_1+c_2-c_3)]\\-
&\frac{1-c}{2}log_2(1-c)-\frac{1+c}{2}log_2(1+c)
\end{split}
\end{equation}
where $c$ is defined as $c:=max\{|c_1|,|c_2|,|c_3|\}$, $\emph{I}(\rho)$ is quantum mutual information, and $\emph{C}(\rho)$ is measurement mutual information. The basis entropy of this state is
\begin{equation}
\emph{BE}=S(\sum_jB_j\rho B_j)-S(\rho)
\end{equation}
where $\{B_j=V\Pi_kV^{\dagger}\otimes V\Pi_lV^{\dagger}:j=0,1,2,3;k=0,1;l=0,1\}$, $\{\Pi_k=|k\rangle\langle k|:k=0,1\}$, $V\in U(2)$. And $V$ can be written as $V=t\textbf{I}+i\overrightarrow{y}\overrightarrow{\sigma}$, with $t\in R$, $\overrightarrow{y}=(y_1,y_2,y_3)\in R^3$, and $t^2+y_1^2+y_2^2+y_3^2=1$.
Which means
\begin{equation}
\begin{split}
 \rho_{pm}=&\sum_jB_j\rho B_j\\
 =&\frac14(\textbf{I}+\sum_i\sum_k\sum_lc_iV\Pi_kV^{\dagger}\sigma_iV\Pi_kV^{\dagger}\otimes V\Pi_lV^{\dagger}\sigma_iV\Pi_lV^{\dagger})
\end{split}
\end{equation}
After calculation, we get
\begin{equation}
  \rho_{pm}=\frac14(\textbf{I}+(c_1z_1^2+c_2z_2^2+c_3z_3^2)V\sigma_3V^{\dagger}\otimes V\sigma_3V^{\dagger})
\end{equation}
where $z_1=2(-ty_2+y_1y_3),z_2=2(ty_1+y_2y_3),z_3=t^2+y_3^2-y_1^2-y_2^2$. The eigenvalues of this $\rho_{pm}$ is
\begin{equation}
  \begin{split}
    \mu_1=&\frac{1-c_1z_1^2-c_2z_2^2-c_3z_3^2}{4}\\
    \mu_2=&\frac{1-c_1z_1^2-c_2z_2^2-c_3z_3^2}{4}\\
    \mu_3=&\frac{1+c_1z_1^2+c_2z_2^2+c_3z_3^2}{4}\\
    \mu_4=&\frac{1+c_1z_1^2+c_2z_2^2+c_3z_3^2}{4}
  \end{split}
\end{equation}
while the eigenvalues of $\rho$ is
\begin{equation}
  \begin{split}
    \lambda_1=\frac{1-c_1-c_2-c_3}{4}\\
    \lambda_2=\frac{1-c_1+c_2+c_3}{4}\\
    \lambda_3=\frac{1+c_1-c_2+c_3}{4}\\
    \lambda_4=\frac{1+c_1+c_2-c_3}{4}\\
  \end{split}
\end{equation}
So the basis entropy will be
\begin{equation}
  \begin{split}
    S(\rho_{pm})-S(\rho)=&\frac14[(1-c_1-c_2-c_3)log_2(1-c_1-c_2-c_3)\\+
&(1-c_1+c_2+c_3)log_2(1-c_1+c_2+c_3)\\+
&(1+c_1-c_2+c_3)log_2(1+c_1-c_2+c_3)\\+
&(1+c_1+c_2-c_3)log_2(1+c_1+c_2-c_3)]\\-
&\frac{1-c_1z_1^2-c_2z_2^2-c_3z_3^2}{2}log_2(1-c)-\frac{1+c_1z_1^2+c_2z_2^2+c_3z_3^2}{2}log_2(1+c)
  \end{split}
\end{equation}
Make $c=max\{|c_1|,|c_2|,|c_3|\}$ and notice that $z_1^2+z_2^2+z_3^2=1$, we get the minimum basis entropy:
\begin{equation}
\begin{split}
  S(\rho_{pm})-S(\rho)=&\frac14[(1-c_1-c_2-c_3)log_2(1-c_1-c_2-c_3)\\+
&(1-c_1+c_2+c_3)log_2(1-c_1+c_2+c_3)\\+
&(1+c_1-c_2+c_3)log_2(1+c_1-c_2+c_3)\\+
&(1+c_1+c_2-c_3)log_2(1+c_1+c_2-c_3)]\\-
&\frac{1-c}{2}log_2(1-c)-\frac{1+c}{2}log_2(1+c)
\end{split}
\end{equation}
which is exactly the quantum discord between subsystems $A$ and $B$.\\
\indent Quantum discord between subsystems in state $\rho=(\textbf{I}+\sum_{i=1}^3c_i\sigma_i\otimes\sigma_i)/4$ is a symmetric quantity, but for others joint states, quantum discord may not be a symmetric quantity. For example, for state like
\begin{equation}
  \rho=\frac14
  \begin{pmatrix}
    2&0&0&0\\
    0&0&0&0\\
    0&0&1&1\\
    0&0&1&1
  \end{pmatrix}
\end{equation}
\indent The quantum discord $\delta(A:B)_{\{\Pi_i^B\}}$ is nonzero, while $\delta(B:A)_{\{\Pi_i^A\}}$ is zero. Although the quantum discord of this system is not symmetric, we can prove that the minimum basis entropy of this state is nonzero, which means theorem 2 still holds. One thing we should note here is that, when the quantum discord is not symmetric, the minimum basis entropy is not exactly the same as the non-zero quantum discord. Take the above state (C.10) as an example, its quantum discord $\delta(A:B)_{\{\Pi_i^B\}}=0.1887$, the measurement projectors are
\begin{equation}
  B_0=\frac12
  \begin{pmatrix}
    1&0\\
    0&0
  \end{pmatrix}
  ,B_1=\frac12
  \begin{pmatrix}
    0&0\\
    0&1
  \end{pmatrix}
\end{equation}
while the minimum basis entropy of the joint state is
\begin{equation}
\begin{split}
  BE_{min}=&S[\frac14
  \begin{pmatrix}
    2&0&0&0\\
    0&0&0&0\\
    0&0&1&0\\
    0&0&0&1
  \end{pmatrix}
  ]-S[\frac14
  \begin{pmatrix}
    2&0&0&0\\
    0&0&0&0\\
    0&0&1&1\\
    0&0&1&1
  \end{pmatrix}
  ]\\
  =&1.5-1\\
  =&0.5\neq0.2896
  \end{split}
\end{equation}
So, the minimum basis entropy of the joint system could be used as a detector of quantum discord, it would be inappropriate to use it to quantify quantum discord in some states.


\begin{thebibliography}{99}
\bibitem{note2}For a joint system $\rho^{AB}$, the quantum discord of this system comes from $S(A|\{\Pi_B\})-S(A|B)$, the conditional entropy $S(A|B)$ is increased by measureing $B$ in orthogonal projectors $\{\Pi_B\}$
\end{thebibliography}
\end{document}